\documentclass[]{article}  

\usepackage{mhd209}
\usepackage{epsfig}
\usepackage{array}
\usepackage{amsmath}

\newcommand\Rm{\mbox{\textit{$R_m$}}}   
\newcommand{\ovec}[1]{{\mbox{\boldmath $#1$}}}
\newcommand{\bb}{\ovec{b}}
\newcommand{\bB}{\ovec{B}}
\newcommand{\bC}{\ovec{C}}
\newcommand{\be}{\ovec{e}}
\newcommand{\bE}{\ovec{\cal{E}}}

\newcommand{\bff}{\ovec{f}}

\newcommand{\bu}{\ovec{u}}
\newcommand{\bU}{\ovec{U}}
\newcommand{\bx}{\ovec{x}}
\newcommand{\bX}{\ovec{X}}

\newcommand{\bmB}{\overline{\ovec{B}}}
\newcommand{\bmu}{\overline{\ovec{u}}}

\title{Harmonic and subharmonic solutions\\ 
of the Roberts dynamo problem.\\
Application to the Karlsruhe experiment}

\author{F. Plunian$^1$  and K.-H. R\"adler$^2$} 

\institute{$^1$ Laboratoires des Ecoulements G\'{e}ophysiques et Industriels\\ 
B.P. 53, 38041 Grenoble Cedex 9, France\\ 
Franck.Plunian@hmg.inpg.fr\\
$^2$Astrophysikalisches Institut Potsdam\\ 
An der Sternwarte 16, D-14482, Potsdam, Germany}

\begin{document}

\maketitle

\begin{abstract}
Two different approaches to the Roberts dynamo problem are considered.
Firstly, the equations governing the magnetic field are specified 
to both harmonic and subharmonic solutions and reduced 
to matrix eigenvalue problems, which are solved numerically.
Secondly, a mean magnetic field is defined by averaging over proper areas,   
corresponding equations are derived, in which the
induction  effect of the flow occurs essentially as an anisotropic
alpha-effect,   
 and they are solved analytically. 
In order to check the reliability of the statements on the Karlsruhe experiment 
which have been made on the basis of a mean-field theory, 
analogous statements are derived for a rectangular dynamo box
containing 50 Roberts cells, and they are compared with the direct solutions 
of the eigenvalue problem mentioned.
Some shortcomings of the simple mean--field theory are revealed.
\end{abstract}

\section{Introduction.}

\begin{figure}
  \begin{flushright}
  \begin{tabular}{@{\hspace{3cm}}c@{\hspace{0cm}}c@{\hspace{1.5cm}}}
    \raisebox{2.5cm}{$y$}
  \epsfig{file=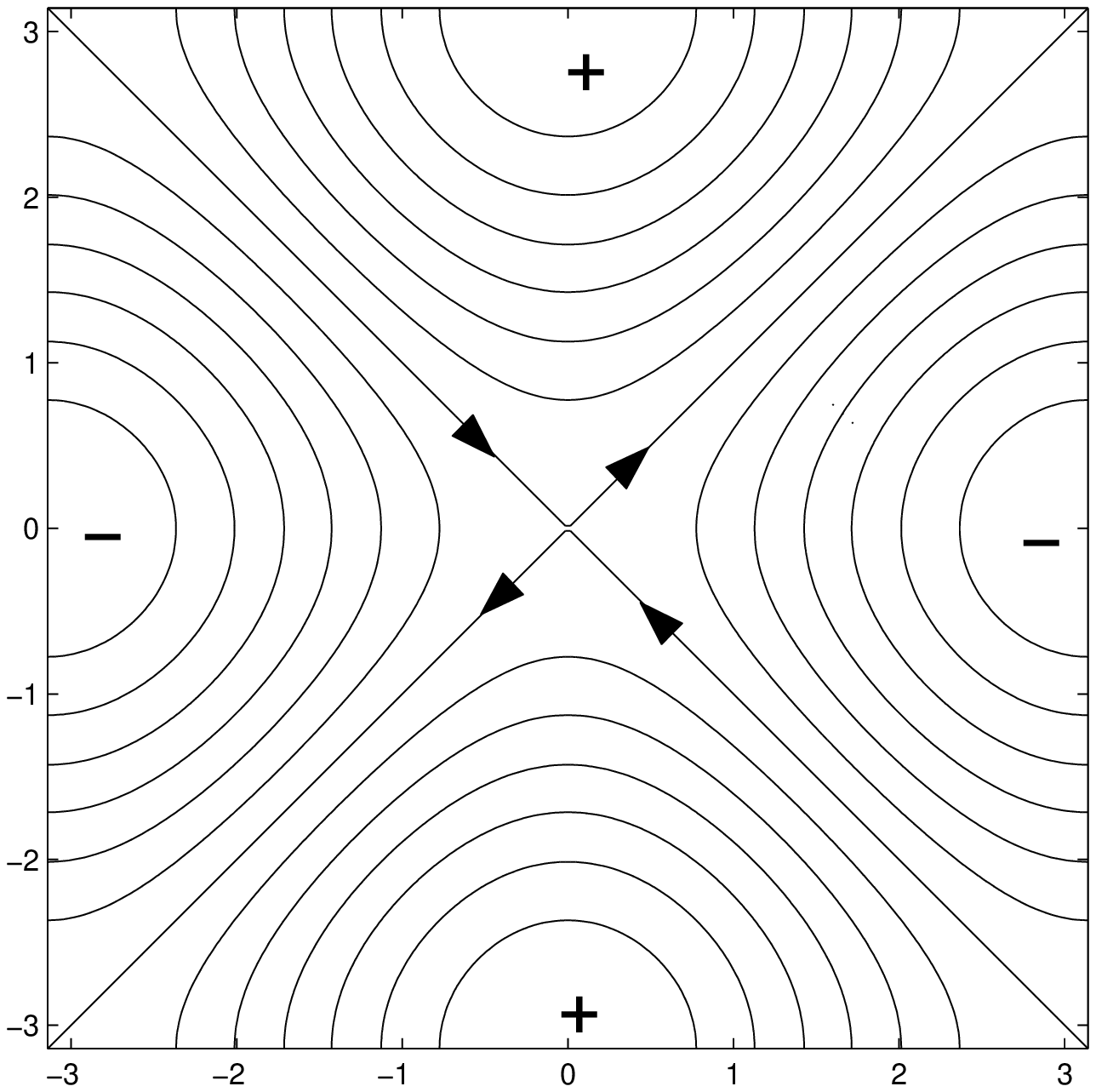,width=0.4\textwidth}
    \\*[0cm]
     & \parbox{0.5\textwidth}{\hspace{-2.5cm} $x$}
    \end{tabular}
\end{flushright}
    \caption{The stream lines of the Roberts flow in the $xy$-plane.
             They coincide with the isolines of the velocity in z-direction.
             Dimensionless coordinates  $x = X / L_U$ and $y = Y / L_U$ are used.}
\label{fig:roberts}
\end{figure}

The dynamo model proposed by Roberts 1972 \cite{roberts72} has been chosen 
as the starting point for an experimental demonstration of homogeneous fluid dynamo
at the Forschungszentrum Karlsruhe 
\cite{busseetal96,muelleretal00,muelleretal02,raedleretal96,raedleretal02,
raedleretal02b,stieglitzetal00,stieglitzetal02,tilgner97}.
The fluid velocity field considered  by Roberts which is of particular interest 
in this context is given by
\begin{equation}
\bU = U \big(\sin(Y/L_U), \, \cos(X/L_U), \, \chi (\cos(X/L_U) - \cos(Y/L_U)) \big) \, .
\label{new1}
\end{equation}
Here a Cartesian coordinate system $(X, Y, Z)$ is used.  
The flow pattern is sketched in Fig. \ref{fig:roberts}.
$L_U$ is the length of the diagonal of a cell in the $XY$--plane
and the parameter $\chi$, which is a constant, determines the $Z$--component
of the flow and so the helicity of the velocity field.
Roberts has demonstrated that a flow of this kind is capable of dynamo action. 
He investigated, however, only magnetic fields which show the same periodicity
in $X$ and $Y$ as the flow pattern. 
These fields, which we call here ``harmonic fields", possess parts which do not depend 
on $X$ and $Y$ but only on $Z$ or, in other words, they have infinite wave lengths
in the $X$ and $Y$ directions.
As Roberts himself pointed out the considered flow allows also 
non--decaying magnetic fields with finite wave lengths in all directions.
For a particular case such fields were investigated  by Tilgner and Busse 
\cite{tilgneretal95}, 
who called them ``subharmonic",
and in a more general frame by Plunian and R\"adler 
\cite{plunianetal02}.
Despite the finite dimensions of the Karlsruhe experimental device 
many estimates concerning excitation conditions etc. have been made on the basis
of findings about harmonic magnetic fields. 
It is, however, of high interest to compare these results with such derived  
from results on subharmonic fields.

In this paper we start with the basic equations of the Roberts dynamo problem 
and some general consequences (Section \ref{Robdyn}), 
present some findings on its harmonic solutions 
(Section \ref{Harm}) 
and explain a mean--field approach to the dynamo problem on that level 
(Section \ref{Mfappr}).            
After that we turn to subharmonic solutions and give some results for them   
(Section \ref{sec: sub model}).
We then deal with the Karlsruhe experiment, 
derive in the framework of a mean--field approach
and under simplifying assumptions on the dynamo module 
an excitation condition and compare it with a corresponding result 
of the subharmonic analysis
(Section \ref{sec:ApplKar}).  
Finally summarize the main consequences of our findings
(Section \ref{sec:conclusion}). 

\section{The Roberts dynamo.}
\label{Robdyn}

To discuss the Roberts dynamo problem in some detail we consider 
the induction equation governing the magnetic field $\bB$, 
assuming that it applies in all infinite space. 
We use its dimensionless form  
\begin{equation}
\frac{\partial \bB}{\partial t} =
\nabla \times(\bu \times \bB) + \Rm^{-1} \nabla^{2}\bB,
\;\;\;\; \nabla \cdot \bB = 0 
\label{inductiondim}
\end{equation}
with 
\begin{equation}
\bu = ( \sin y, \sin x, \chi(\cos x -\cos y) ) \, .
\label{robflow}
\end{equation}
Instead of $\bX = (X, Y, Z)$ we have introduced here dimensionless coordinates 
$\bx = (x, y, z)$ defined by $\bx = \bX / L_U$, 
instead of $\bU$ the dimensionless velocity $\bu$ defined by $\bu = \bU / U$,
and we measure the time in units of $L_U / U$.
Further $\Rm$ is the magnetic Reynolds number defined by
\begin{equation}  
\Rm = \frac{UL_U}{\eta}  
\label{rm}  
\end{equation}   
with $\eta$ being the magnetic diffusivity of the fluid.

For a steady flow as envisaged here we may expect solutions $\bB$ 
varying like $\exp(pt)$ in time,
where the real part of $p$ is the dimensionless growth rate.
In this case an eigenvalue problem for $\bB$ 
with the eigenvalue parameter $p$ occurs. 
Furthermore, since the flow is $z$-independent, 
$\bB$ can be assumed to possess the form 
\begin{equation}
\bB = \Re\{\bb(x,y,k) \exp(pt + \mbox{i}kz)\} \, ,
\label{eq:bB}
\end{equation}
where $\bb (x,y,k)$ is a complex vector field independent of $z$,
and $k$ a dimensionless wave number with respect to the $z$-direction.
When inserting (\ref{eq:bB}) into (\ref{inductiondim}) we find
\begin{equation}
p\bb + (\bu \cdot \nabla)\bb =
(\bb\cdot \nabla)\bu - \mbox{i} k u_z\bb +
 \Rm^{-1}(\nabla^{2} - k^2) \bb
\label{eq:bharm}
\end{equation}
\begin{equation}
\nabla \cdot \bb + \mbox{i} k b_z=0.
\label{eq:divbharm}
\end{equation}
The $x$ and $y$ components of (\ref{eq:bharm}) are equations for $b_x$ and $b_y$ 
which do not contain $b_z$.
They constitute the mentioned eigenvalue problem. 
After solving it, we can calculate
$b_z$ from (\ref{eq:divbharm})
without any integration.

We may easily conclude from (\ref{eq:bharm}) that the results for $p$ 
for any $\chi$ can be inferred from those for $\chi=1$ with the help of
the relation 
\begin{equation}
p(k,\Rm, \chi)=p(k\chi,\Rm, 1)+k^2\Rm^{-1}(\chi^2-1).
\label{ki}
\end{equation}
As long as we deal with direct solutions of the Roberts dynamo problem 
(up to Section \ref{sec: sub model})
we therefore restrict our attention to the solutions with $\chi=1$.
In the discussion of the Karlsruhe dynamo experiment (Section \ref{sec:ApplKar}) 
we admit also other values of $\chi$.

\section{Harmonic solutions.}
\label{Harm}

As mentioned above, Roberts solved the relevant equations only for magnetic fields 
with the same periods in $x$ and $y$ as the flow pattern.
We consider first this case only, in which we speak of ``harmonic solutions". 
Then $\bb (x,y,k)$ must have the same periodicity in $x$ and $y$ as the flow pattern.

\begin{figure}   
\begin{flushright}
  \begin{tabular}{@{\hspace{2cm}}c@{\hspace{0cm}}c@{\hspace{1.5cm}}}
    \raisebox{2.5cm}{$p$}
 \epsfig{file=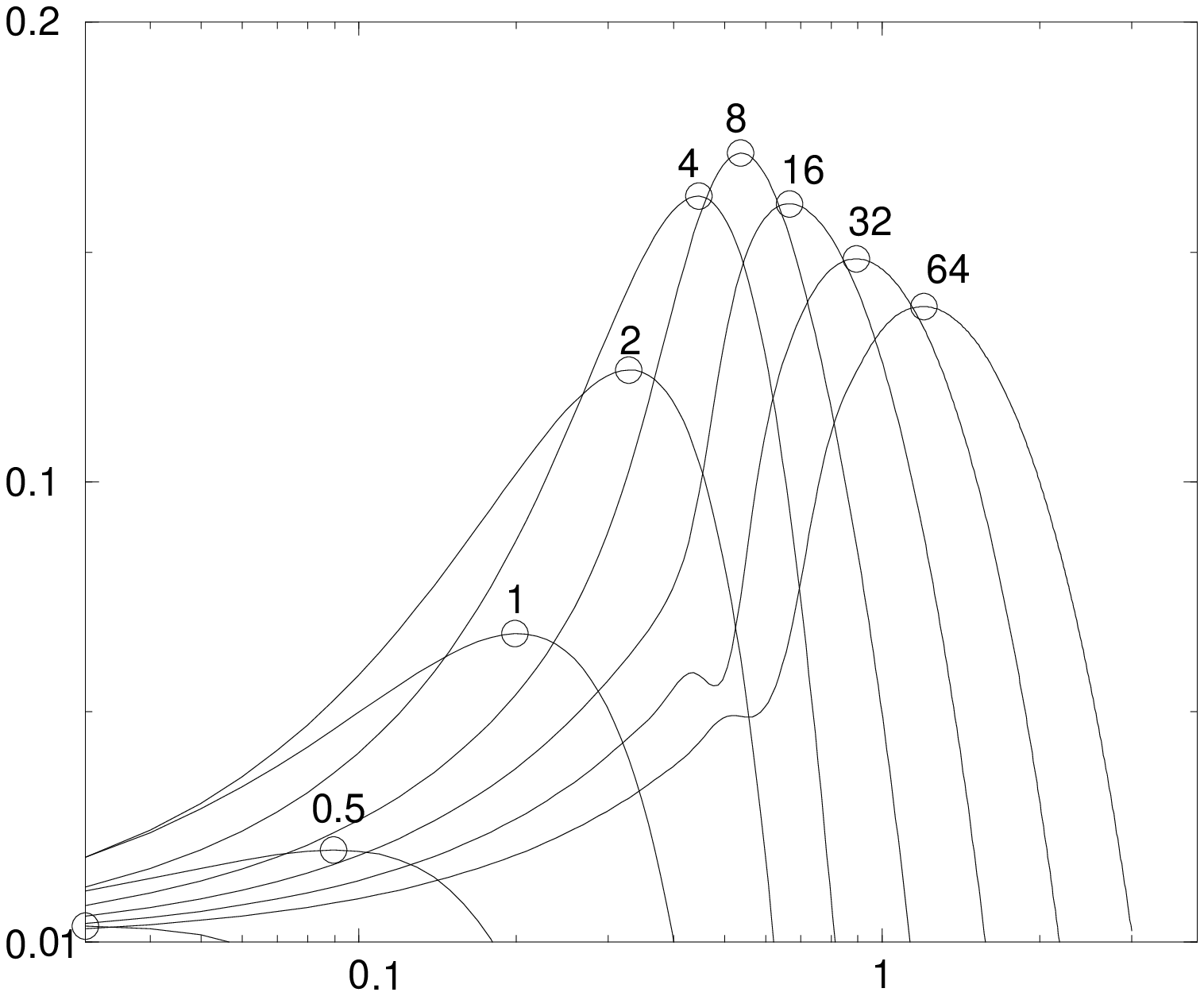,width=0.6\textwidth}
    \\*[0cm]
     & \parbox{0.5\textwidth}{\hspace{-2.5cm} $k$}
    \end{tabular}
\end{flushright}
    \caption{The growth rate $p$ versus $k$ for various $\Rm$ 
    (given by the labels at the curves) and $\chi=1$. 
    For each given $\Rm$ the circle corresponds to the maximum growth rate $p_{\max}$ 
    occurring at $k_{\max}$.}
\label{fig:p}
\end{figure}

Solving the eigenvalue problem defined by (\ref{eq:bharm}) Roberts found 
growth rates, that is real parts of  $p$, in its dependence on $k$ 
for values of $\Rm$ up to 64 as shown in Fig. \ref{fig:p}.
The imaginary parts of $p$ proved to be equal to zero (numerically always close to zero), 
attesting that the dynamo instability is an absolute one.
This implies that the magnetic field geometry is stationary 
while the intensity in general varies in time. 

Soward \cite{soward87} has shown that in the limit of large $\Rm$ 
the order of the maximum of the dimensionless growth rate, $p_{\max}$,
is given by $p_{\max} = O(\ln(\ln\Rm)/\ln\Rm)$. 
It occurs at a wave number $k_{\max}$ 
for which $k_{\max}= O(\Rm^{1/2}(\ln\Rm)^{-1/2})$.
That is, $p_{\max} \to 0$ as $\Rm \to \infty$. 
Thus the Roberts dynamo proves to be a slow one.
This applies not only in the context of harmonic solutions, 
for it turns out that other solutions never grow faster 
than the fastest of the harmonic ones.

Results for finite $\Rm$ obtained in our calculations are given in Fig. \ref{fig:p} 
and in Table \ref{tab:maxpkrm}; see also \cite{soward89,soward90}.
Note that the quantities $\Gamma_1 = p_{\max} \ln\Rm / \ln(\ln\Rm)$ 
and $\Gamma_2 = k_{\max} (\ln\Rm)^{1/2}/\Rm^{1/2}$ given 
in Table \ref{tab:maxpkrm} approach constant values as $\Rm$ grows 
and thus illustrate the mentioned asymptotic laws.    

\begin{table}
  \begin{center}
  \begin{tabular}{lcccccccccc}
  \hline
       $\Rm$           & 2    & 4    & 8    & 16   & 32   & 64   & 128  & 256  & 512   \\
       \hline
       $p_{\max}$      & 0.12 & 0.16 & 0.17 & 0.16 & 0.15 & 0.14 & 0.13 & 0.12 & 0.11  \\
       $k_{\max}$      & 0.33 & 0.45 & 0.54 & 0.66 & 0.89 & 1.2  & 1.62 & 2.18  & 2.87 \\
       $\Gamma_1$      & -0.24& 0.69 & 0.49 & 0.44 & 0.41 & 0.40 & 0.40 & 0.39 & 0.37  \\
       $\Gamma_2$      & 0.19 & 0.26 & 0.28 & 0.28 & 0.29 & 0.31 & 0.31 & 0.32 & 0.32\\
  \hline     
  \end{tabular}
  \end{center}
  \caption{Maximum growth rates $p_{max}$, the corresponding wave numbers $k_{max}$ 
  and the quantities $\Gamma_1$ and $\Gamma_2$ for various $\Rm$ and $\chi = 1$.}
\label{tab:maxpkrm}
\end{table}

\section{A mean--field approach.}
\label{Mfappr}

The Roberts dynamo with harmonic magnetic fields $\bB$ can also be understood 
in terms of mean fields. 
For any field $Q$ we define a mean field $\overline{Q}$ by averaging 
over an area of one periodic unit, or four half cells, of the flow pattern 
in the $xy$-plane as depicted in Fig. \ref{fig:roberts}, for example 
\begin{equation}
\overline{Q}(x,y,z,t) = \int_{-\pi}^{+\pi} \int_{-\pi}^{+\pi}
Q(x+ \xi, y+\eta, z, t) d\xi d\eta.
\label{eq:averaging}
\end{equation}
Whereas $\bmB$ is in general non--zero we have $\bmu$ = 0. 

Subjecting the induction equation
(\ref{inductiondim}) to this kind of averaging we obtain
\begin{equation}
\frac{\partial \bmB}{\partial t} = 
\nabla \times \bE + \Rm^{-1} \nabla^{2}\bmB,
\;\;\,\,\,\, \;\;\,\,\,\, \nabla \cdot \bmB = 0,
\label{inducmean}
\end{equation}
where $\bE$ is a mean electromotive force due to the fluid motion,
\begin{equation}
\bE = \overline{\bu \times \bB}.
\label{eq:electromotive force}
\end{equation}
We admit here no other magnetic fields $\bB$ than harmonic ones in the above sense.
Then $\bmB$ and $\bE$ depend no longer on $x$ and $y$ but only on $z$ and $t$.

Let us further use Fourier representations for $\bB$, $\bmB$ and $\bE$ according to 
\begin{equation}
Q (x,y,z,t) = \int \hat{Q} (x,y,k,t) e^{\mbox{i}kz} \mbox{d} k \, ,  
\label{kh01} 
\end{equation}
where the integral is over $- \infty < k < \infty$. 
The corresponding representation of $\bB$ clearly includes the ansatz (\ref{eq:bB}).
$\hat{\bB}$ depends on $x$, $y$, $k$ and $t$, 
but $\hat{\bmB}$ and $\hat{\bE}$ depend only on $k$ and $t$.
The requirements that $\bB$, $\bmB$ and $\bE$ are real lead to  
$\hat{\bB}^* (x,y,k,t) = \hat{\bB} (x,y,-k,t)$ 
and analogous relations for $\hat{\bmB}$ and $\hat{\bE}$. 

According to (\ref{eq:electromotive force}) we have 
\begin{equation}
\hat{\bE} (k, t) = \overline{\bu (x, y) \times \hat{\bB} (x, y, k, t)} \, .
\label{kh03} 
\end{equation}
With standard reasoning of mean-field theory 
we conclude that $\bE$ is linear and homogeneous in $\bmB$.
For the sake of simplicity we assume that $\bE$ at a given time 
depends only on $\bmB$ at the same time, that is,
ignore any dependence on $\bmB$ at earlier times. 
We therefore write
\begin{equation}
{\hat{\cal{E}}}_i (k, t)  
    =  {\hat{\alpha}}_{ij} (k) {\hat{\overline{B}}}_j (k, t) \, ,
\label{kh05} 
\end{equation}
where ${\hat{\alpha}}_{ij}$ is a complex tensor determined by the fluid flow. 
Analogous to $\bE$ and $\bmB$ it has to satisfy 
${\hat{\alpha}}_{ij}^* (k) = {\hat{\alpha}}_{ij} (-k)$. 
From the symmetry properties of the $\bu$--field we conclude further 
that the relation (\ref{kh05}) remains its validity if both $\bE$ and $\bmB$ 
are simultaneously subject to a $90^{o}$ rotation about the $z$--axis, 
and that $\bE$ has no $z$--component 
and does not depend on the $z$--component of $\bmB$.  
The first fact means that the tensor ${\hat{\alpha}}_{ij}$ is axisymmetric 
with respect to the $z$--axis, that is, can only be a linear combination 
of $\delta_{ij}$, $e_i e_j$ and $\epsilon_{ijl} e_l$, 
where $\delta_{ij}$ is the Kronecker tensor, 
$\epsilon_{ijl}$ the Levi--Civita tensor
and $\be$ the unit vector in $z$--direction.
The second fact implies that $\delta_{ij}$ and $e_i e_j$ occur only 
in the combination $\delta_{ij} - e_i e_j$.
Thus we may write  
\begin{equation}
{\hat{\alpha}}_{ij} (k) = - {\hat{\alpha}}_1 (k) (\delta_{ij} - e_i e_j)
    + \mbox{i} k {\hat{\alpha}}_2 (k) \epsilon_{ijl} e_l
\label{kh07} 
\end{equation}
with two complex functions ${\hat{\alpha}}_1 (k)$ and ${\hat{\alpha}}_2 (k)$.
The signs in this relation and the factor $\mbox{i} k$ of the last term
have here to be considered as arbitrary  
but will prove to be useful in the following.  

\begin{figure}
  \begin{flushright}
  \begin{tabular}{@{\hspace{2cm}}c@{\hspace{0cm}}c@{\hspace{1.5cm}}}
    \raisebox{2.5cm}{$\hat{\alpha}$}
 \epsfig{file=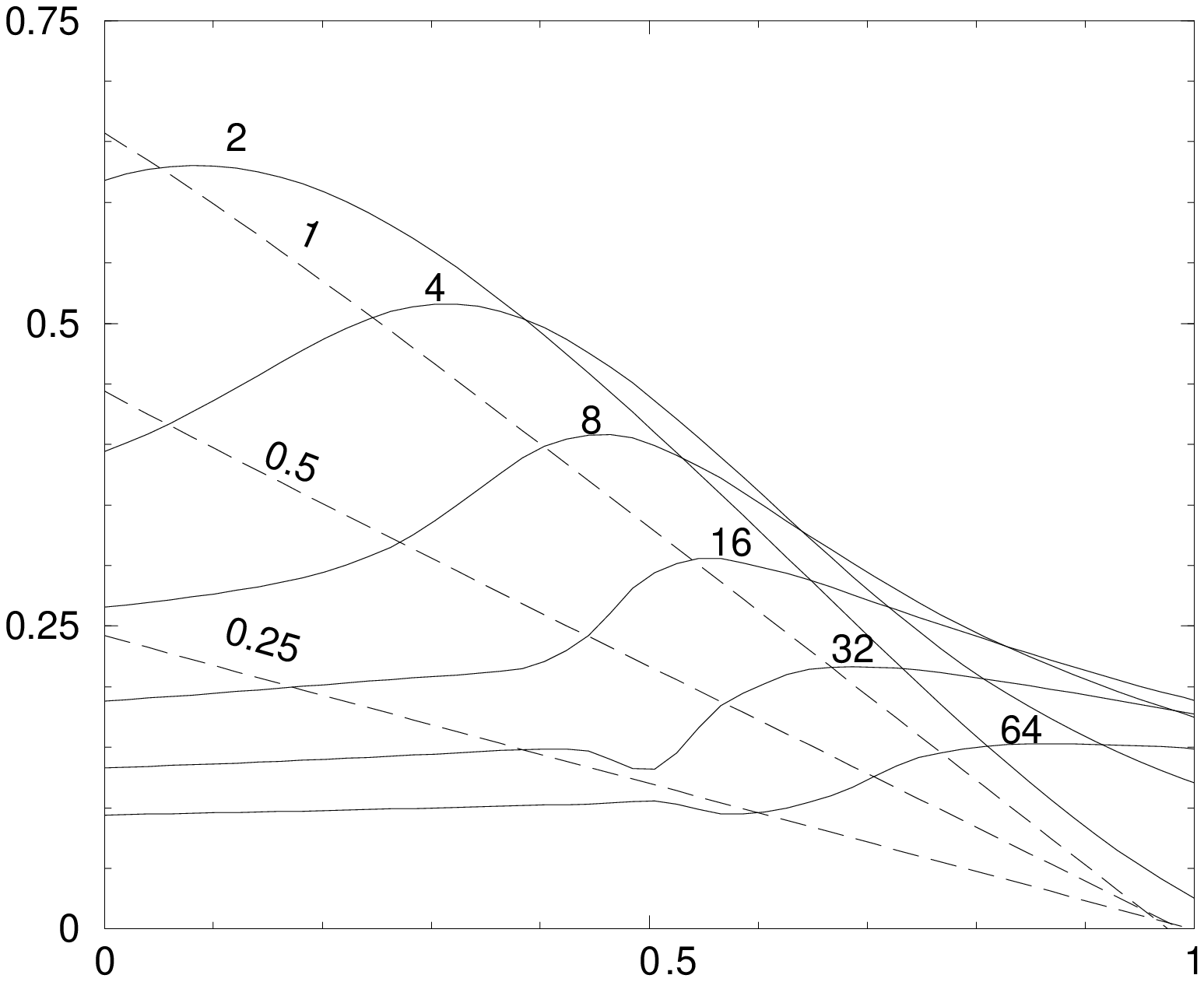,width=0.6\textwidth}
    \\*[0cm]
     & \parbox{0.5\textwidth}{\hspace{-2.5cm} $k$}
    \end{tabular}
\end{flushright}
    \caption{The coefficient $\hat{\alpha}$ versus $k$ for various $\Rm$ 
    (given by the labels on the curves) and $\chi=1$.}
\label{fig:alpha}
\end{figure}

As can be easily seen from the equations (\ref{inducmean}) 
and (\ref{eq:electromotive force}) 
their solutions must have the form 
\begin{equation}
\bmB = \Re \{\bC \exp (pt + \mbox{i}kz) \} 
\label{kh08}
\end{equation}
with a complex vector $\bC$ lying in the $xy$--plane.
If we use (\ref{kh05}) and (\ref{kh07}) we find that 
\begin{equation} 
C_x = \pm \mbox{i} C_y
   \label{CxCy} 
\end{equation} 
and 
\begin{equation}
p + \Rm^{-1} k^2 = {\hat{\alpha}} (k)  k  \, , \quad
{\hat{\alpha}} (k) = \pm {\hat{\alpha}}_1 (k) - {\hat{\alpha}}_2 (k) k \,
\label{simple1}
\end{equation}
The two signs in (\ref{CxCy}) and (\ref{simple1}) indicate the existence of two
classes of solutions and had been already identified by Roberts
\cite{roberts72}.  Relations of the same kind have also been found by Soward
\cite{soward87}   (who used the notations $\alpha_C$ and $\alpha_D$  
instead of -${\hat{\alpha}}_1$ and ${\hat{\alpha}}_2$).  
We can calculate ${\hat{\alpha}}(k)$ with the help of the harmonic solutions  
of the induction equation (\ref{inductiondim}) obtained with the ansatz (\ref{eq:bB}). 
There are again two classes of these harmonic solutions,  
which correspond to the two signs in (\ref{CxCy}) and (\ref{simple1}). 
The two classes lead to two different functions ${\hat{\alpha}}(k)$. 
In agreement with the finding that $p$ is real also ${\hat{\alpha}}$ proves to be real.  
Among the two function ${\hat{\alpha}}(k)$ only the largest one (the upper
sign in (\ref{CxCy}) and (\ref{simple1})) has been retained as it corresponds
to the largest growthrate $p$ (see also \cite{plunianetal02}). Fig.
\ref{fig:alpha} shows the dependence of ${\hat{\alpha}}$ on $k$
 for various
$\Rm$ and $\chi=1$. 
 The values of ${\hat{\alpha}}$ for arbitrary $\chi$ can
be inferred from the values 
 for $\chi=1$ with the help of the relation 
\begin{equation}
 {\hat{\alpha}}(k,\Rm,\chi)=\chi {\hat{\alpha}}(k \chi,\Rm,
1) \, , 
 \label{alphaki}
\end{equation}
which follows from (\ref{ki}) and (\ref{simple1}).

Since ${\hat{\alpha}}$ is real we may assume that also
${\hat{\alpha}}_1$ and ${\hat{\alpha}}_2$ are real.
Moreover we may then conclude from ${\hat{\alpha}}_{ij}^* (k) = {\hat{\alpha}}_{ij} (-k)$
that both are even in k, 
that is ${\hat{\alpha}}_1 (k) = {\hat{\alpha}}_1 (-k)$ 
and ${\hat{\alpha}}_2 (k) = {\hat{\alpha}}_2 (-k)$.
If we know $\hat{\alpha} (k)$ for positive and negative $k$ we may find
${\hat{\alpha}}_1 (k)$ and ${\hat{\alpha}}_2 (k)$. 

We may conclude with the help of (\ref{kh05}) and (\ref{kh07}) that 
\begin{eqnarray}
\bE (z, t) &=&  
   - \int {\hat{\alpha}}_1 (k) \big( \hat{\bmB} (k, t) 
   - (\be \cdot \hat{\bmB} (k, t)) \be \big) 
   \exp(\mbox{i}kz) \mbox{d}k 
\nonumber \\
    && \quad - \be \times \frac{\partial}{\partial z} 
    \int {\hat{\alpha}}_2 (k) \hat{\bmB} (k, t) 
    \exp(\mbox{i}kz) \mbox{d}k \, . 
\label{kh09}
\end{eqnarray}
This is equivalent to
\begin{eqnarray}
\bE (z, t) &=&  - \frac{1}{2 \pi} 
    \int \alpha_1 (\zeta) \big( \bmB (z + \zeta, t) 
    - (\be \cdot \bmB (z + \zeta, t)) \be \big) 
    \mbox{d}\zeta 
\nonumber \\
    && \quad - \frac{1}{2 \pi} \be \times \frac{\partial}{\partial z} 
    \int \alpha_2 (\zeta) \bmB (z + \zeta, t) \mbox{d} \zeta 
\label{kh11}
\end{eqnarray}
with
\begin{equation}
\alpha_1 (\zeta) = \int {\hat{\alpha}}_1 (k) \exp( \mbox{i} k \zeta ) \mbox{d} k \, , \quad
    \alpha_2 (\zeta) = \int {\hat{\alpha}}_2 (k) \exp( \mbox{i} k \zeta ) \mbox{d} k \, .
\label{kh13} 
\end{equation}
The integrals are again over $- \infty < k < \infty$ 
or over $- \infty < \zeta < \infty$. 
Note that both $\alpha_1$ and $\alpha_2$ are even in $\zeta$.

Let us now expand $\hat{\alpha}_{ij} (k)$ in a Taylor series.
>From (\ref{kh07}) we have
\begin{equation}
{\hat{\alpha}}_{ij} (k) 
    = - {\hat{\alpha}}_1 (0) ( \delta_{ij} - e_i e_j )  
    + \mbox{i} k {\hat{\alpha}}_2 (0) \epsilon_{ijl} e_l  + \cdots \, .
\label{eq:expansion}
\end{equation} 
The second relation (\ref{simple1}) together with the symmetry properties 
of ${\hat{\alpha}}_1$ and ${\hat{\alpha}}_2$ leads to 
\begin{equation}
{\hat{\alpha}}_1 (0) = {\hat{\alpha}} (0) \, , \quad    
   {\hat{\alpha}}_2 (0) = - \frac{\partial \hat{\alpha}}{\partial k} (0) \, .
\label{kh15}
\end{equation} 
The corresponding expansion of $\bE$ reads
\begin{equation}
\bE = - \alpha_\perp \big( \bmB - (\be \cdot \bmB) \be \big)
     - \beta \be \times \frac{\partial \bmB}{\partial z} + \cdots \, ,
\label{kh17}
\end{equation}
where 
\begin{equation} 
\alpha_\perp =  \hat{\alpha}(0) \quad \, , \quad 
    \beta = - \frac{\partial \hat{\alpha}}{\partial k} (0) \, .
\label{kh19}
\end{equation} 
The same result can be derived from (\ref{kh11}) by expanding $\bmB (z + \zeta, t)$
with respect to $\zeta$.
The first term on the right--hand side of (\ref{kh17}) describes 
the anisotropic $\alpha$--effect.
Since here $\be \times \frac{\partial \bmB}{\partial z} = \nabla \times \bmB$
it seems at the first glance reasonable to consider the second term 
as a contribution to a mean--field diffusivity.
By a reason mentioned later in Section \ref{mfexp}, however, this interpretation 
is not compelling.

Values of $\alpha_\perp$ and $\beta$ for various $\Rm$ are given in Table
\ref{tab:alpha}.
 Remarkably enough $\beta$ is negative as soon as $\Rm$
exceeds a value between 1 and 2.
 This means that then the corresponding term
in (\ref{kh17}) 
 supports the dynamo action of the flow. 
For the limit of large $\Rm$ it was shown that $\alpha_\perp = O(\Rm^{-1/2})$ 
\cite{childress79,raedleretal98}.
This is illustrated by the values of the quantity $\Gamma_3 = \alpha_\perp \Rm^{1/2}$
given in Table \ref{tab:alpha}.  
\begin{table}
  \begin{center}
  \begin{tabular}{lcccccccccc}
  \hline
       $\Rm$              			& 0.25 & 0.5  & 1    & 2    & 4    & 8    & 
16   & 32   & 64   \\
\hline
       $\alpha_\perp$   			& 0.24 & 0.44 & 0.66 & 0.62 & 0.39 & 0.27 & 
0.19 & 0.13 & 0.09 \\
       $\beta$                                & 0.25 & 0.46 & 0.54 & -0.3 & -0.4 & 
-0.12&-0.07 & -0.05& -0.03\\      
       $\Gamma_3$          			& 0.12 & 0.32 & 0.66 & 0.87 & 0.79 & 0.75 & 
0.75 & 0.75 & 0.75 \\
\hline
  \end{tabular}
  \end{center}
  \caption{The coefficients $\alpha_\perp$ and $\beta$
  of the expansion (\ref{kh17}) 
  and the quantity $\Gamma_3$
  for various $\Rm$ and $\chi = 1$.}
  \label{tab:alpha}
\end{table}

\section{Subharmonic solutions.}
\label{sec: sub model}

In order to check a simple mean-field theory of the Karlsruhe dynamo experiment 
we are interested in solutions $\bB$ of the induction equation (\ref{inductiondim})
with period lengths exceeding those of the flow pattern,
which we call ``subharmonic solutions". 
We consider the case in which the period lengths of $\bB$ are larger 
by an integer factor $N$ than those of the flow pattern.   
As mentioned above, this problem has already been investigated by
Tilgner \& Busse \cite{tilgneretal95} for a few special values of $N$
and later by Plunian and R\"adler \cite{plunianetal02}.
                                          
We focus our attention again on the induction equation (\ref{inductiondim})
governing the magnetic field $\bB$ in all space.
We use again (\ref{eq:bB}) but consider $\bb$ no longer as a field 
with the same periodicity in $x$ and $y$ as the flow pattern.
Instead we put $\bb = \tilde{\bb}(x,y,f_x,f_y,k) \exp(\mbox{i}( f_x x + f_y y))$,
where $\tilde{\bb}$ has now the same periodicity as the flow pattern
and $f_x$ and $f_y$ are subharmonic wave numbers in the $x$ and $y$ directions.      
In that sense we look for solutions of the induction equation of the form
\begin{equation}
\bB = \Re \left( \tilde{\bb}(x,y,\bff) \exp( pt + \mbox{i} \bff \cdot \bx ) \right) \, .
\label{eq:Bdef}
\end{equation}
with $\tilde{\bb}(x,y,\bff)$ being a complex periodic vector field  
with the same period length in $x$ and $y$ direction as the flow pattern,
$\bff$ the  real vector $(f_x, f_y, k)$, 
and $p$ again a complex quantity;   
for more details see \cite{plunianetal02}.
We restrict our attention here to the case $f_x = f_y = f = 1/N$. 
Then the period lengths of the magnetic field $\bB$ are just $N$ times that of 
the flow pattern. 
The harmonic solutions $\bB$ discussed above correspond 
to the limit $N \to \infty$.

\begin{figure}   \begin{flushright}
  \begin{tabular}{@{\hspace{2cm}}c@{\hspace{0cm}}c@{\hspace{1.5cm}}}
    \raisebox{2.5cm}{$\Rm$}
 \epsfig{file=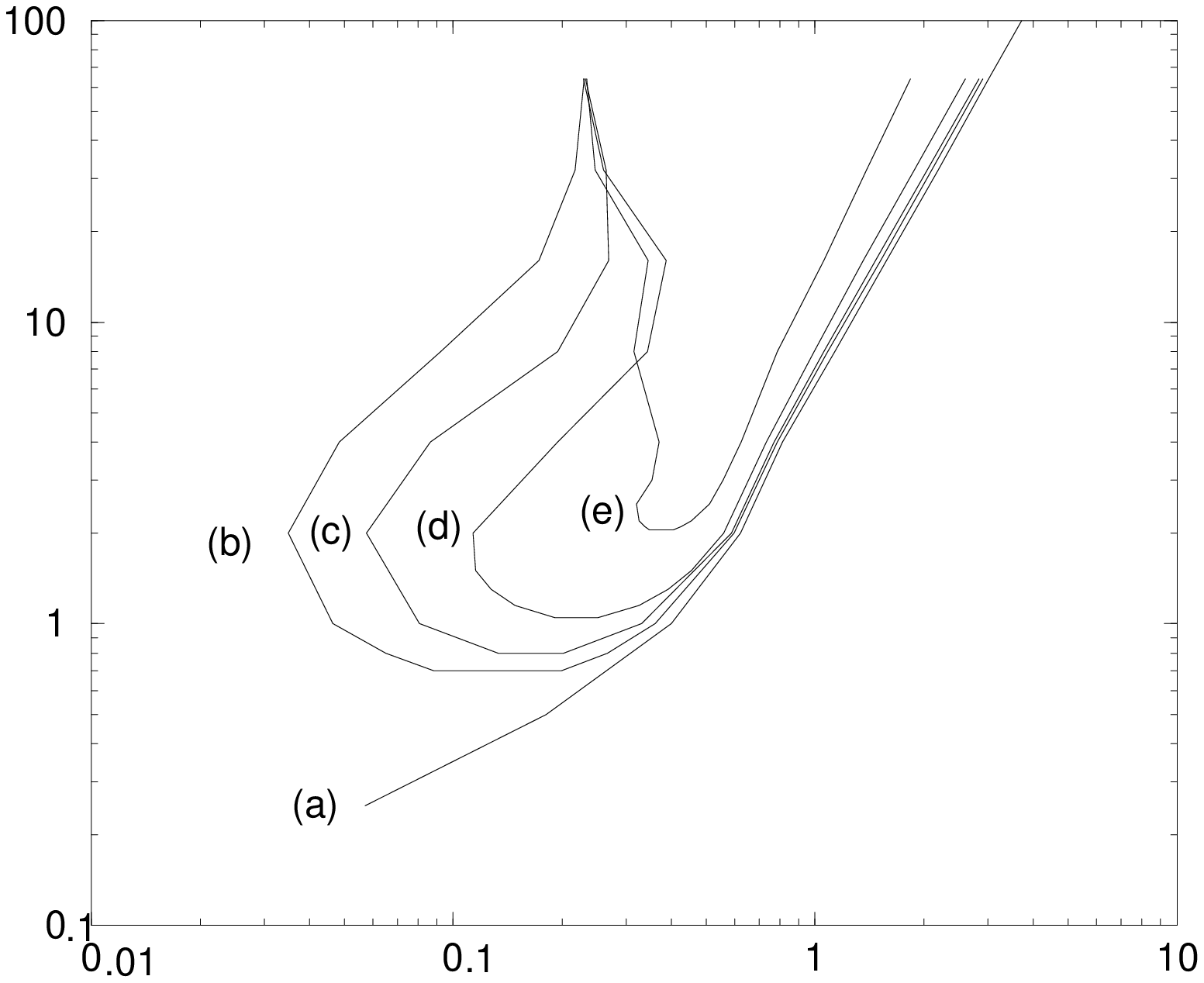,width=0.6\textwidth}
    \\*[0cm]
     & \parbox{0.5\textwidth}{\hspace{-2.5cm} $k$}
    \end{tabular}
\end{flushright}
    \caption{Marginal values of $\Rm$ versus $k$ for $f_x=f_y=f$ and
    $\chi=1$. The curve (a) belongs to  $f=0$, (b) to $f=1/10$, (c) to $f=1/8$,
    (d) to $f=1/6$, (e) to $f=1/4$.}  \label{fig:marginal}
\end{figure}

Inserting (\ref{eq:Bdef}) in (\ref{inductiondim}) we find
\begin{equation}
p\tilde{\bb} + (\bu \cdot \nabla)\tilde{\bb} =
(\tilde{\bb} \cdot \nabla)\bu -i(\bff \cdot \bu)\tilde{\bb} +
 \Rm^{-1}( \nabla^{2}\tilde{\bb} +2i(\bff \cdot \nabla)\tilde{\bb} - \bff^2\tilde{\bb} ),
 \label{eq:b2}
\end{equation}
\begin{equation}
\nabla \cdot \tilde{\bb} + i\tilde{\bb} \cdot \bff =0 \,.
\label{eq:divb}
\end{equation}
The system (\ref{eq:b2}) 
defines an eigenvalue problem with $p$ being the eigenvalue
parameter\footnote{These equations have already been derived by Roberts
(1972) though used only with $f_x=f_y=0$ in his numerical
calculations.}. 
It has been solved numerically. 
Marginal values of $\Rm$ versus $k$ for $f_x=f_y=f$ are shown in Fig. \ref{fig:marginal} 
for different values of $f$ and again $\chi=1$. 
For $f \ne 0$ there are both a critical $\Rm$ and a critical $k$ below which
dynamo action is not possible. 
In general $p$ is no longer real, that is, we have no longer stationary but moving
field structures.  
We point out that relation (\ref{ki}) allows us the calculation of $p$
for arbitrary $\chi$ from that for $\chi = 1$.

\section{Applications to the Karlsruhe experiment.}
\label{sec:ApplKar}

\subsection{The Karlsruhe ``dynamo module".}
\label{subsec:Karlsruhe}
 
The essential piece of the Karlsruhe dynamo experiment
is the ``dynamo module", a cylindrical container with both radius and height
somewhat less than 1 m, through which liquid sodium is driven by external pumps 
\cite{muelleretal00,muelleretal02,stieglitzetal00,stieglitzetal02}.   
By means of a system of channels with conducting walls,
constituting  52 ``spin generators", a helical motion is organized.  
The flow pattern is similar to that defined by (\ref{new1}).
The 52 spin generators correspond to 26 periodic units of the flow pattern
such as the one shown in Fig. \ref{fig:roberts}.
The arrangement of the pumps allows to vary the parameters $U$, or $\Rm$, and $\chi$ 
independently from each other.

\subsection{A simple mean--field theory of the experiment.}
\label{mfexp}
In order to give an estimate for the self--excitation condition 
of the experimental device a simple mean--field theory has been developed
\cite{raedleretal97a,raedleretal97b,raedleretal98,raedleretal02,raedleretal02b,
raedleretal02c}.
The mean magnetic field $\bmB$ defined as above is assumed to satisfy 
the equations (\ref{inducmean}) and (\ref{eq:electromotive force}) inside the dynamo module 
and to continue in some way in outer space.  
Of course, in this context $\bmB$ can no longer be independent on $x$ and $y$,
and $\bE$ can no longer have the simple forms (\ref{kh09}), (\ref{kh11}) or (\ref{kh17}).
Relying on some traditional concept it was assumed that the variations of $\bmB$
in space are sufficiently weak so that $\bE$ in a given point can be represented 
by $\bmB$ and its first spatial derivatives in this point.  
Together with the symmetry properties of the Roberts flow this leads to 
\begin{eqnarray}
\bE &=& -\alpha_{\perp}(\bmB - (\be \cdot \bmB) \be)
    - \beta_{\perp}\nabla \times \bmB 
    - (\beta_{\parallel} - \beta_{\perp})(\be \cdot (\nabla \times \bmB)) \be
\nonumber \\
    && - \beta_3 \be \times (\nabla (\be \cdot \bmB) + (\be \cdot \nabla) \bmB) \, ,
\label{meanemf}
\end{eqnarray}
where $\alpha_{\perp},\beta_{\perp},\beta_{\parallel}$ and $\beta_3$
are constants depending on $\Rm$ and $\chi$, 
and $\be$ is again the unit vector in $z$-direction 
\cite{raedleretal96,raedleretal02b}.
As in (\ref{kh17}) the term with $\alpha_\perp$ describes the anisotropic $\alpha$--effect
acting in the $xy$--plane only.
The terms with $\beta_{\perp}$ and $\beta_{\parallel}$ can be interpreted 
by introducing an anisotropic mean-field diffusivity 
different from the molecular magnetic diffusivity.
Finally the term with $\beta_3$ describes a part of $\bE$ 
depending on derivatives of $\bmB$ which cannot be expressed by $\nabla \times \bmB$
and therefore not be interpreted as a contribution to a modified diffusivity.   
Several results have been derived on the dependence of $\alpha_\perp$ 
on the fluid flow
\cite{raedleretal96,raedleretal97a,raedleretal97b,raedleretal98,raedleretal02,
raedleretal02b}, 
and also such on $\beta_\perp$, $\beta_\parallel$ and $\beta_3$  
\cite{raedleretal96}.

For a field $\bmB$ not depending on $x$ and $y$ and having no $z$--component
we have $\nabla \times \bmB = \be \times \frac{\partial \bmB}{\partial z}$, 
and the last three terms on the right--hand side of (\ref{meanemf})
can be written in the form 
$ - (\beta_\perp + \beta_3) \be \times \frac{\partial \bmB}{\partial z}$.
As to be expected, in this special case 
the structures of (\ref{kh17}) and (\ref{meanemf}) coincide, 
and we have $\beta = \beta_\perp + \beta_3$.
Our above remark on the $\beta_3$--term in (\ref{meanemf})
explains why the interpretation of the $\beta$--term in (\ref{kh17})
as a contribution to a mean-field diffusivity is not compelling.
 
The assumption on small variations of $\bmB$ in space means in particular 
that $\bmB$ does not change markedly across a spin generator.
In that sense the usage of (\ref{meanemf}) in a theory of the dynamo module 
can only be justified for a very large number of spin generators within the module.
Quite a few solutions of the equation (\ref{inducmean}) for $\bmB$, 
applied to the dynamo module, with $\bE$ according to (\ref{meanemf}) 
and various boundary conditions have been calculated 
\cite{raedleretal98,raedleretal02,raedleretal02b}. 
In most cases, however, no other contribution to $\bE$ than the $\alpha$--effect,
that is, only the first term on the right--hand side of (\ref{meanemf}) 
was taken into account.
Contributions with higher than first derivatives of $\bmB$ have never been considered.
By these and other reasons a check of the results of the simple mean--field theory 
on a way that avoids the mentioned shortcomings seems very desirable.

\subsection{Comparison of results of mean-field approach 
and subharmonic analysis.}
\label{sec:results}

For this purpose we deal now with a very simple model of the dynamo module.
We consider no longer a cylindrical module but instead a rectangular ``dynamo box"
with a quadratic basis area in the $xy$--plane 
and denote the edge lengths of the box in this plane by $L$
and its hight by $H$.
Thinking of the shape of the real dynamo module we put $L / H = 2$.
We will study the excitation condition for a mean magnetic field $\bmB$ 
which satisfies the equation (\ref{inducmean}) 
and the relation (\ref{eq:electromotive force}) in all space 
and is periodic in $x$ and $y$ with the period length $2 L$ 
and in $z$ with the period length $2 H$.
This periodicity means that the dynamo box contains just a ``half wave" of the field $\bmB$.
For the sake of simplicity we use (\ref{eq:electromotive force}) 
in its reduced form containing no other induction effect than the $\alpha$--effect. 
We will then compare this excitation condition with that for a subharmonic $\bB$--field 
whose longest wave lengths show the same periodicity, 
that is, which fits in the same sense to the dynamo box.
In this context we put $N = 10$ so that an area of $(2 L)^2$ in the $xy$--plane 
contains just 100 period units, that is, 200 cells of the flow pattern, 
consequently the basis area of the dynamo box 50 cells, 
which have to be compared with the 52 spin generators in the real dynamo module. 
$N = 10$ means $f = 1 / 10$, and with $L / H = 2$ we arrive at $k = 1 / 5$.  

Instead of a realistic boundary condition for the dynamo module we use here in fact
the condition of periodic continuation of the magnetic fields
both on the mean--field and the subharmonic level. 
Such a condition might be in general problematic 
but seems acceptable for the comparison which we have in mind.

From Fig. \ref{fig:alpha} we see that the value of $\hat{\alpha}$ for $k = 1/5$ can, 
except for small $\Rm$,
not be inferred from $\hat{\alpha}(0)$ 
and $\frac{\partial \hat{\alpha}}{\partial k}(0)$ only.
This suggests that there will be discrepancies between the excitation conditions 
obtained with a mean--field theory which ignores contributions to $\bE$ 
with higher than first-order spatial derivatives and those derived 
from the subharmonic analysis.

As already explained we assume for our mean--field consideration 
that equation (\ref{inducmean}) and the reduced form of (\ref{meanemf}), 
that is,
\begin{equation}
\frac{\partial \bmB}{\partial t} =
   - \nabla \times \big( \alpha_{\perp}(\bmB - (\be \cdot \bmB) \be) \big) 
   + \Rm^{-1} \nabla^{2}\bmB, \;\;\,\,\,\, \nabla \cdot \bmB = 0 \, ,
\label{inducmean2}
\end{equation}
apply in all space with constant $\alpha_{\perp}$.
We may represent $\bmB$ as a sum of a
poloidal and a toroidal part,
\begin{equation}
\bmB = -\nabla \times (\be \times \nabla S) - \be \times \nabla T \, ,
\label{polotoro}
\end{equation}
with two defining scalars $S$ and $T$. Inserting this in
(\ref{inducmean2}) and dropping unimportant constants we find 
\begin{eqnarray}
\Rm^{-1} \Delta S - \alpha_{\perp}T -\frac{\partial S}{\partial t} & = & 0 \nonumber \\
\Rm^{-1} \Delta T + \alpha_{\perp} \frac{\partial^2 S}{\partial z^2} -\frac{\partial
T}{\partial t} & = & 0 \, .
\label{system}
\end{eqnarray}
The special periodic solution of (\ref{inducmean2}) which we are looking for
is obtained with the ansatz
\begin{eqnarray}
  S &=& S_0 \;\; cos\left(f x \right)
            cos\left(f y \right)
            cos\left(k z \right)
	    \exp(pt)
\nonumber \\
  T &=& (T_0/S_0)S \, ,
\label{ST}
\end{eqnarray}
where $S_0$ and $T_0$ are constants, $f$ and $k$ the parameters specified above 
and $p$, which will prove to be real, is again the growth rate. 
When inserting this in (\ref{system}) we arrive 
at two linear homogeneous equations for $S_0$ and $T_0$. 
The requirement that they allow non-trivial solutions leads to
\begin{equation}
p = - \frac{2 f^2 + k^2}{\Rm} \pm |\alpha_{\perp}| k \, .
\label{dispersion}
\end{equation}
Growing $\bmB$ are possible in the case of the upper sign of the last term   
if $|\alpha_{\perp}|$ is sufficiently large. 
The excitation condition reads
\begin{equation}
|\alpha_{\perp}| \geq  \frac{2 f^2 + k^2}{\Rm k} \, .
\label{excitation}
\end{equation}

In the representations of results on $\alpha_\perp$ 
on which we now rely    
\cite{raedleretal96,raedleretal97a,raedleretal97b,raedleretal98} 
the latter is given in its original dimension so that it corresponds 
$U \alpha_\perp$ with our dimensionless $\alpha_\perp$.
Furthermore these results are given in terms of the two magnetic Reynolds numbers 
$\Rm_{\parallel}$ and $\Rm_{\perp}$ for the flow in the $xy$-plane 
and in $z$-direction, respectively. 
These are connected with our $\Rm$ and $\chi$ by
\begin{equation}
\Rm_{\parallel} = \frac{8\sqrt{2}}{\pi} \chi \Rm,\;\;\;\;\;\;
\Rm_{\perp} = 2 \Rm.
\label{rmdef}
\end{equation}

\begin{figure}
  \begin{flushright}
  \begin{tabular}{@{}l@{\hspace{-1cm}}c@{\hspace{1.5cm}}}
    \raisebox{5cm}{$\Rm_{\parallel}^*$}
  \epsfig{file=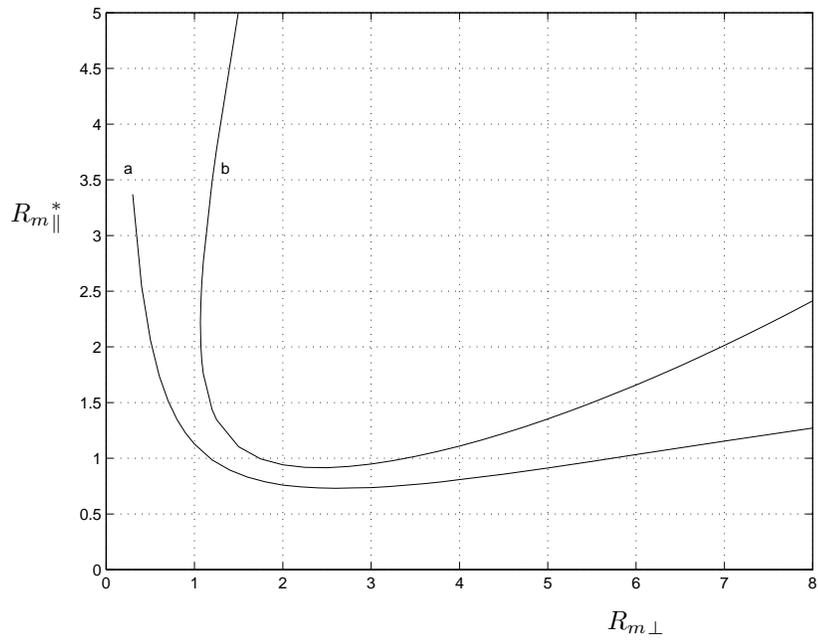,width=0.8\textwidth}
    \\*[0cm]
     & \parbox{0.5\textwidth}{\hspace{-2cm} $\Rm_{\perp}$}
    \end{tabular}
\end{flushright}
\caption{Marginal values of $\Rm_{\parallel}^*$ versus $\Rm_{\perp}$.
Curve (a) results from a mean-field calculation [6].
Curve (b) is derived from the subharmonic analysis with
$f=1/10$ and $k=1/5$.}
\label{fig:f/k const}
\end{figure}

The marginal states of the dynamo, in which $\bmB$ neither grows nor decays,
are given by pairs of $\Rm_{\perp}$ and $\Rm_{\parallel}$, 
or by the corresponding neutral curve in the $\Rm_{\perp} \Rm_{\parallel}$--diagram.
We may represent the result for arbitrary $f$ and $k$ by using 
the modified magnetic Reynolds number $\Rm_{\parallel}^*$ defined by 
\begin{equation}
\Rm_{\parallel}^* = \frac{\pi}{16 \sqrt{2}} \Rm_{\parallel}
\frac{k}{2 f^2  + k^2}
\end{equation}
instead of $\Rm_{\parallel}$.
Note that $\Rm_{\parallel}^*$ is no longer determined by $\bU$ alone but also by $\bB$.
Fig. \ref{fig:f/k const} shows a $\Rm_{\perp} \Rm_{\parallel}^*$--diagram 
in which  curve (a) gives just the result of our mean--field calculation.
Clearly dynamo action requires that $\Rm_{\parallel}^*$  exceeds a critical value.
It appears, however, to be possible for any $\Rm_{\perp}$ if only $\Rm_{\parallel}^*$
is sufficiently large.

Let us now compare this result with that for a corresponding subharmonic solution $\bB$ 
of the induction equation. 
In Fig. \ref{fig:f/k const} the curve (b) is the neutral one for the subharmonic solution 
with the values of $f$ and $k$ specified above.
Clearly dynamo action requires now not only 
that $\Rm_{\parallel}^*$  exceeds a critical value 
but also that $\Rm_{\perp}$ lies above such a value.
In addition for each given $\Rm_{\perp}$ allowing dynamo action  
the marginal value of $\Rm_{\parallel}^*$ derived in the subharmonic approach 
is higher than that concluded from the mean--field approach.
In the range of $\Rm_{\perp}$ between 1.2 and 2, 
which corresponds to the actual situation in the Karlsruhe experiment,
the deviation is larger than 20 \%.   
Of course it will become smaller in a comparison with a mean--field model
which involves also the induction effects connected with first derivatives 
of $\bmB$ indicated in (\ref{meanemf}); see \cite{raedleretal96}.
But even then the mean--field approach underestimates the requirements 
for self--excitation.   

\section{Conclusions.} 
\label{sec:conclusion} 

We have dealt with several aspects of the Roberts dynamo problem
and derived some results which are of interest for the Karlsruhe dynamo experiment.
Although a rectangular dynamo box was considered, there are good reasons 
to assume that the main conclusions apply as well to the real experimental device 
with a cylindrical dynamo module.
In the framework of the simple mean--field theory of the experiment 
self--excitation seems possible for arbitrary values 
of the magnetic Reynolds number $\Rm_\perp$ describing the flow perpendicular
to the axes of the spin--generators 
if only the magnetic Reynolds number $\Rm_\parallel$ for the axial flow 
is sufficiently large.
An analysis based on subharmonic solutions revealed that a dynamo is only possible 
if both $\Rm_\perp$ and $\Rm_\parallel$ exceed critical values.
Apart from this it was found that the simple mean-field theory underestimates 
the excitation condition of the dynamo. 
This discrepancy of the mean-field results with those obtained with subharmonic solutions 
cannot be completely removed by taking into account the effect of the mean--field 
diffusivity.

\bigskip
 
\thanks{The authors are indebted to Dr. M. Rheinhardt for several helpful comments 
on the draft of this paper.}

\bibliographystyle{mahyd}

\newcommand{\noopsort}[1]{}

\end{document}